\documentclass{article}
\usepackage{spconf,amsmath,graphicx,amssymb,color}
\usepackage[ruled]{algorithm2e}
\usepackage[hidelinks]{hyperref}
\usepackage{subcaption}
\usepackage{bm}
\usepackage[OT1]{fontenc}
\DeclareMathOperator*{\argmin}{argmin}
\newcommand*{\argminl}{\argmin\limits}
\usepackage{booktabs}

\title{Visual Privacy Protection via Mapping Distortion}
\name{Yiming Li$^{1,\star}$\thanks{$\star$ indicates equal contribution.} \qquad Peidong Liu$^{1,\star}$\thanks{This work is supported by the National Science Foundation of China under Grant 61771273, the Natural Science Foundation of Zhejiang Province (LSY19A010002), and the R\&D Program of Shenzhen (JCYJ20180508152204044). Corresponding author: Yong Jiang. } \qquad Yong Jiang$^{1,2}$ \qquad Shu-Tao Xia$^{1,2}$}

            \address{$^{1}$Tsinghua Shenzhen International Graduate School, Tsinghua University\\
            $^{2}$PCL Research Center of Networks and Communications, Peng Cheng Laboratory \\
                  \{li-ym18, lpd19\}@mails.tsinghua.edu.cn;
                \{jiangy, xiast\}@sz.tsinghua.edu.cn}

\begin{document}
%
\maketitle
\begin{abstract}
Privacy protection is an important research area, which is especially critical in this big data era. To a large extent, the privacy of visual classification data is mainly in the mapping between the image and its corresponding label, since this relation provides a great amount of information and can be used in other scenarios. In this paper, we propose the mapping distortion based protection (MDP) and its augmentation-based extension (AugMDP) to protect the data privacy by modifying the original dataset. In the modified dataset generated by MDP, the image and its label are not consistent ($e.g.$, a cat-like image is labeled as the dog), whereas the DNNs trained on it can still achieve good performance on benign testing set. As such, this method can protect privacy when the dataset is leaked.  Extensive experiments are conducted, which verify the effectiveness and feasibility of our method. The code for reproducing main results is available at \url{https://github.com/PerdonLiu/Visual-Privacy-Protection-via-Mapping-Distortion}.
\end{abstract}
\section{Introduction}
\label{sec:intro}

Deep learning, especially deep neural networks (DNNs), have been successfully adopted in many fields, such as object detection \cite{zhu2019feature,wang2019,ahmed2020density}, super-resolution \cite{zhang2019ranksrgan,dai2019second,soh2020meta}, and visual tracking \cite{Dai_2019_CVPR,voigtlaender2020siam,liao2020pg}. A large amount of training data is one of the key factors in the success of DNNs. While the massive amount of data dramatically improves the performance of the DNNs, the collection of data from millions of users also raises serious privacy issues. For example, the collected data has the risk of being leaked, which harms the privacy of users. Accordingly, how to protect the privacy of data is of great significance.

\vspace{0.2em}
In this paper, we focus on protecting data privacy in image classification tasks. To a large extent, the privacy of the tasks is mainly in the ground-truth mapping between the input image and its corresponding label, since this relation provides a significant amount of information and can be used in other scenarios. To the best of our acknowledge, there is only one research, $i.e.$, k-anonymity \cite{k-anonymity}, in this research area. Specifically, k-anonymity hides the mapping by guaranteeing that individuals who are the subjects of the data can not be re-identified while the private data remains practically available. However, this method focused on the field-structured data, which can not be adopted in protecting visual data.

\vspace{0.2em}
To address this problem, we propose a mapping distortion based protection (MDP) and its augmentation-based extension (AugMDP). Our approaches aim at exploring a new possible way to protect the visual privacy by distorting the ground-truth mapping between the image and its label. In other words, for a specific image in the modified dataset generated by the proposed method, its provided label does not match what the image truly is. In this way, we can still protect the privacy when the dataset is leaked, under the condition that the hacker has no prior knowledge of the ground-truth mapping. Besides, models trained with the modified dataset can still achieve good performance on the benign testing set, which guarantees the utility of the generated dataset.

\vspace{0.2em}
Specifically, the modified image is generated by minimizing the distance with a random image with the provided label in the feature space. The mechanism behind MDP is that the DNNs utilize lots of unstable yet useful features such as texture, as discussed in \cite{texture_1, texture_2, ilyas2019adversarial}. It is precisely by hiding the information of the given label in the modified image that DNNs can learn the ground-truth relation even when the provided mapping seems to be distorted. Besides, MDP has two extra latent advantages, including \textbf{(1)} The hackers can hardly realize the incorrectness of the dataset since the perturbation in the modified image is invisible. \textbf{(2)} The leakage can be detected if the specific distorted mapping appears.

\begin{figure}[ht]
\centering
\includegraphics[width=0.42\textwidth]{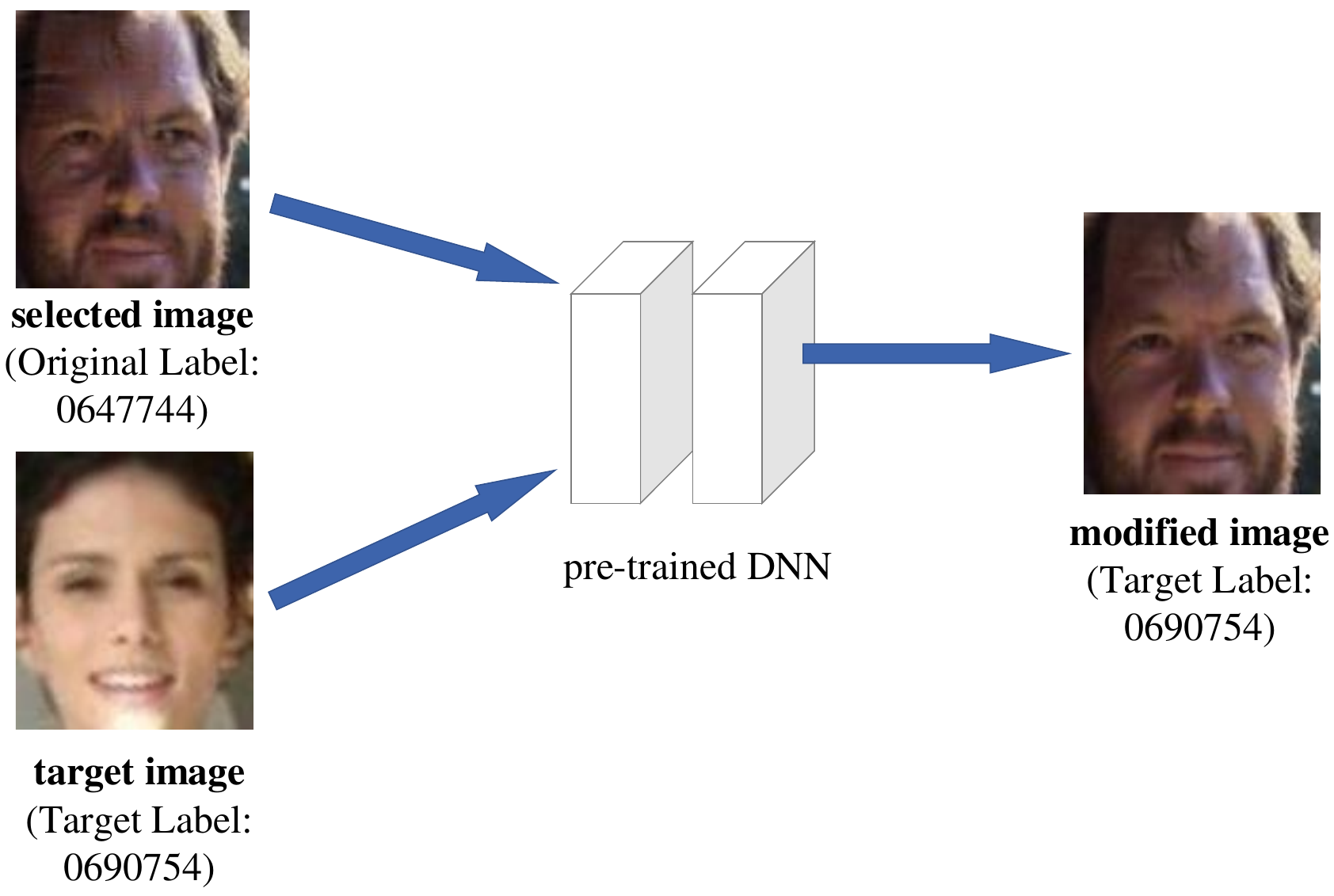}
\caption{The illustration of selected image, target image, modified image, and their corresponding label. The modified image looks similar to its corresponding selected image, whereas its label is the label of the corresponding target image (which is different from that of the selected image). Accordingly, the ground-truth mapping between image and label of the original dataset is hidden and therefore is protected. }
\vspace{-0.8em}
\label{fig:interpret}
\end{figure}

\vspace{0.2em}
The main contributions of our work can be summarized as follows: \textbf{(1)} We explore a novel angle of privacy protection by modifying the ground-truth relation between image and its corresponding label. \textbf{(2)} We provide a simple yet effective method, the MDP, and its augmentation-based extension (AugMDP) for privacy protection. \textbf{(3)} Extensive experiments are conducted, which verifies the feasibility and effectiveness of our privacy protection method.

\section{The Proposed Method}
\label{sec:pagestyle}

\subsection{Preliminary}
Suppose $\mathcal{D} = \{(\bm{x}_i, y_i)\}_{i=1}^{N}$ is the original dataset needed to be protected, where $N$ is the size of the dataset and the input-label pair $(\bm{x}, y) \in \mathcal{X} \times \mathcal{Y}$.  Before the detailed discussion, we first define some key concepts, as visualized in Fig. \ref{fig:interpret}.

\begin{itemize}
    \item \textbf{Selected Image} $\bm{x}_{selected}$: For a given original dataset $\mathcal{D}$, the selected image $\bm{x}_{selected} \in \{\bm{x}|(\bm{x},y) \in \mathcal{D}\}$. 
    \item \textbf{Original Label} $y_{original}$: The label of the selected image $\bm{x}_{selected}$.
    \item \textbf{Target Image} $\bm{x}_{target}$: The target image is also from the original dataset $\mathcal{D}$, $i.e.,$ $\bm{x}_{target} \in \{\bm{x}|(\bm{x},y) \in \mathcal{D}\}$, while its label $y_{target} \neq y_{original}$.
    \item \textbf{Target Label} $y_{target}$: The label of $\bm{x}_{target}$.
    \item \textbf{Modified Image}: The modified image is visually similar to the selected image. It is obtained through minimizing the distance between the output of the image initialized with the selected image and that of the target image in the middle layer of a given pre-trained DNN. Note that its label is the same as that of the target image, $i.e.,$ $y_{modified} = y_{target}$, therefore the relation within the image-label pair $(\bm{x}_{modified}, y_{modified})$ is distorted to hide the ground-truth relation within $(\bm{x}_{selected}, y_{selected})$.
\end{itemize}

\begin{figure}[tb]
\centering
\includegraphics[width=0.42\textwidth]{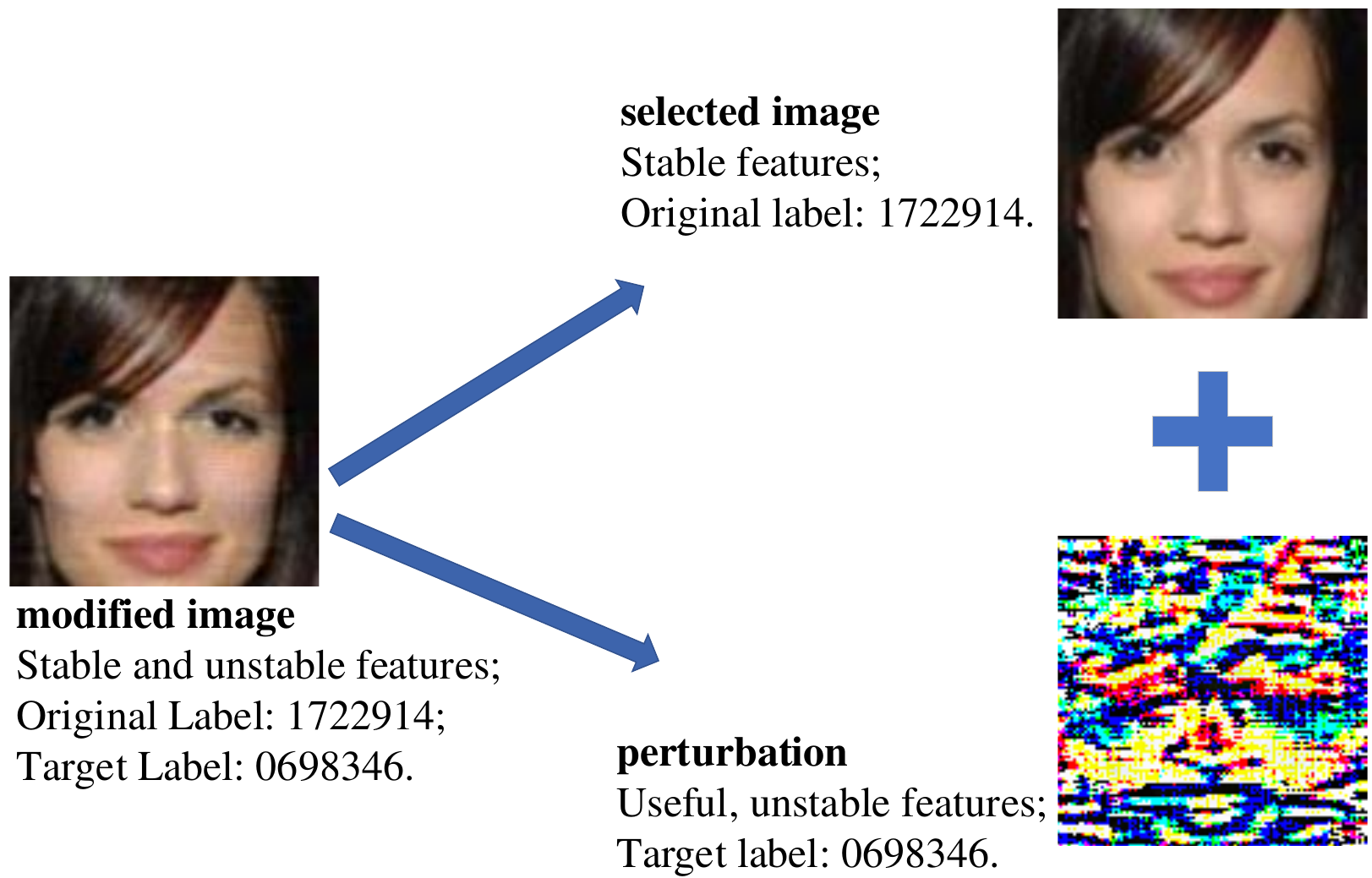}
\caption{The illustration of a modified image. The modified image contains two types of features, including stable features from the selected image and unstable features from the perturbation. The information about ground-truth mapping will be hidden in the unstable features associate with the target label.}
\label{fig:separate}
\vspace{-0.8em}
\end{figure}

As suggested in \cite{no_bugs}, the features used by DNNs can be divided into two categories, including stable and unstable features. Intuitively, stable features ($e.g.,$ the profile \cite{profile_1, profile_2}) are visible and interpretable to humans, while the unstable features, such as the texture \cite{texture_1, texture_2}, are usually invisible. Both stable and unstable features are all useful to the classifier. Since unstable features can be modified easily without being discovered by humans, these features can be utilized to construct a new dataset with distorted mapping, namely the \textbf{modified dataset} to hide to ground-truth mapping for privacy protection. The detailed construction procedure will be discussed in Section \ref{sec:mdp}.

\vspace{-0.6em}
\subsection{Mapping Distortion based Protection}\label{sec:mdp}
As discussed above, hiding the ground-truth mapping between image and its label is critical for privacy protection. Therefore, instead of storing the original dataset directly, we suggest keeping its modified version where the input-label mapping is distorted.

In this paper, we propose a simple yet effective method, dubbed mapping distortion based protection (MDP), for constructing the modified dataset. To obtain useful yet unstable features from the target image, we use a standard pre-trained DNN. Specifically, we first initialize the modified image with the selected image and then minimize the distance between its output and that of the target image in the feature space. We construct the modified training set $\mathcal{D}_{mod}$ via a one-to-one correspondence $\bm{x}_{selected} \rightarrow \bm{x}_{modified}$, where $\bm{x}_{selected}$ is the selected image and $\bm{x}_{modified}$ is the modified image.

\begin{algorithm}[ht]
\small
\KwIn{Original dataset $\mathcal{D}$, Augmentation-related hyper-parameter $T$.}
\KwOut{Augmented dataset $\mathcal{D}_{modAug}$}
Initialize $\mathcal{D}_{modAug} = \{\}$

\For{time in range ($T$)} {
  \For{$(\bm{x}_{target}, y_{target}) \in \mathcal{D}$} {
    Randomly pick a pair $(\bm{x}_{selected}$, $y_{original}) \in \mathcal{D}$. \\
    Calculate $\bm{x}_{modified}$ according to optimization (\ref{equ:modified_sample}).
    \\
    $\mathcal{D}_{modAug} = \mathcal{D}_{modAug} \cup \{(\bm{x}_{modified}, y_{target})\}.$
  }
}
\Return $\mathcal{D}_{modAug}$

\caption{Construction procedure of the augmented dataset.}
\label{algo:algorithm}
\end{algorithm}

Specifically, for every target image $\bm{x}_{target}$ with label $y_{target}$ in $\mathcal{D}$, MDP randomly chooses a selected image $\bm{x}_{selected}$ with original label $y_{original}$ ($y_{original} \neq y_{target}$) in $\mathcal{D}$. After that, the MDP updates $\bm{x}_{modified}$ gradually so that the output of $\bm{x}_{modified}$ and the output of $\bm{x}_{target}$ are similar in the middle layer of the DNN. The update is through the following optimization process:

\vspace{-0.6em}
\begin{equation}
\small
\bm{x}_{modified} = \argminl_{\bm{x} \in [0,1]^D} d\left(f(\bm{x})-f(\bm{x}_{target})\right),
\label{equ:modified_sample}
\end{equation}
where $D$ is the dimension of the features, $f$ is the mapping from input to the output of a certain layer in DNN, and $d(\cdot)$ is a distance metric. We adopt the most widely used $\ell^\infty$ distance, $i.e.$, $d\left(f(\bm{x}_{modified})-f(\bm{x}_{target})\right) = ||f(\bm{x}_{modified})-f(\bm{x}_{target})||_\infty$, for simplicity. More different distance metrics will be discussed in our future work.

Specifically, to solve the optimization problem (\ref{equ:modified_sample}), $\bm{x}$ is initialized with $\bm{x}_{selected}$ and we optimize the problem with the classical projected gradient descent (PGD) method \cite{PGD}. As shown in Fig. \ref{fig:separate}, the modified image is obtained from the combination of the selected image and a small perturbation related to unstable yet useful features. Since those invisible yet useful features contained in the images are still highly predictive, DNNs trained with the modified dataset can still have a good performance on the benign testing test.

\subsection{Augmented Mapping Distortion based Protection}
As mentioned in the previous section, we can store the modified dataset instead of the original one to protect data privacy. However, due to the adverse effects of the incorrect stable features in modified images, training with them will result in a certain decrease in the performance. In this section, we introduce an MDP extension, dubbed augmented MDP (AugMDP), to further enhance the effectiveness of 
our method.

Specifically, in AugMDP, we first construct $T$ different modified datasets $\{\mathcal{D}_{mod}^{(1)}, \cdots, \mathcal{D}_{mod}^{(T)}\}$ through MDP, where $T$ is a positive integer hyper-parameter to control the augmentation size. Then, the augmented dataset is obtained by combining all these datasets, $i.e.$, $\mathcal{D}_{modAug} = \mathcal{D}_{mod}^{(1)} \cup \mathcal{D}_{mod}^{(2)} \cup \dots \cup \mathcal{D}_{mod}^{(T)}$. This augmented method is effective since the extra information carried by the unstable yet useful features in the augmented 
data is conducive to the learning of the model. The construction procedure of $\mathcal{D}_{modAug}$ is shown in Algorithm \ref{algo:algorithm}, and its effectiveness is verified in Section \ref{ssec:augmentation_subsection}.

\begin{figure}[ht]
\centering
\includegraphics[width=0.43\textwidth]{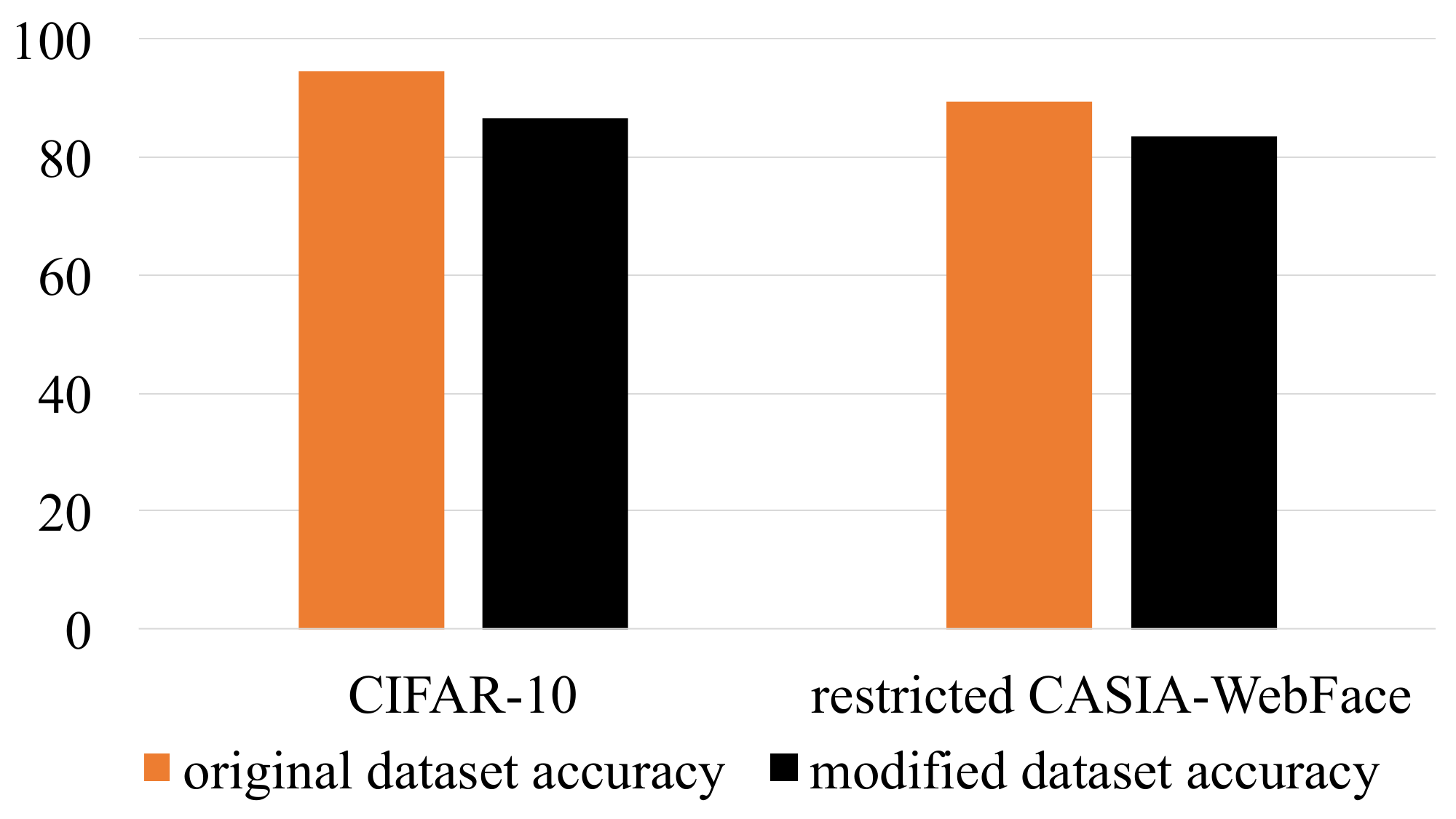}
\caption{Test accuracy on CIFAR-10, restricted CASIA-WebFace, and their corresponding modified dataset.}
\label{fig:unstable_train_effect}
\vspace{-1em}
\end{figure}

\section{Experiments}
\label{sec:majhead}

\subsection{Settings}
The experiments are conducted on the CIFAR-10 \cite{CIFAR10} and (restrict) CASIA-WebFace \cite{CASIA} dataset. Instead of the whole CASIA-WebFace, we only use a subset of the dataset for the consideration of computational complexity. Restricted CASIA-WebFace has 46,492 images with only 1,000 classes, which are randomly chosen from the original dataset. For DNNs, we use ResNet-50 \cite{resnet} and IR-50 with ArcFace (an improved version of the vanilla ResNet-50) \cite{arcface} on CIFAR-10 and restricted CASIA-WebFace datasets, respectively. 

To construct the modified dataset, we perform PGD \cite{PGD} to optimize the objective function under $\ell^{\infty}$ norm, which aims to minimize the distance between the output of the modified image and that of the target image in the penultimate layer of the pre-trained model. Specifically, PGD-100 and PGD-40 with step size 0.1 are used on CIFAR-10 and restricted CASIA-WebFace dataset, respectively. The performance of the trained model is examined on the benign testing set to verify the effectiveness of the modified dataset.

\begin{figure*}[ht]
\begin{center}
\begin{subfigure}{1\textwidth}
    \centering
    \includegraphics[width=0.6\textwidth]{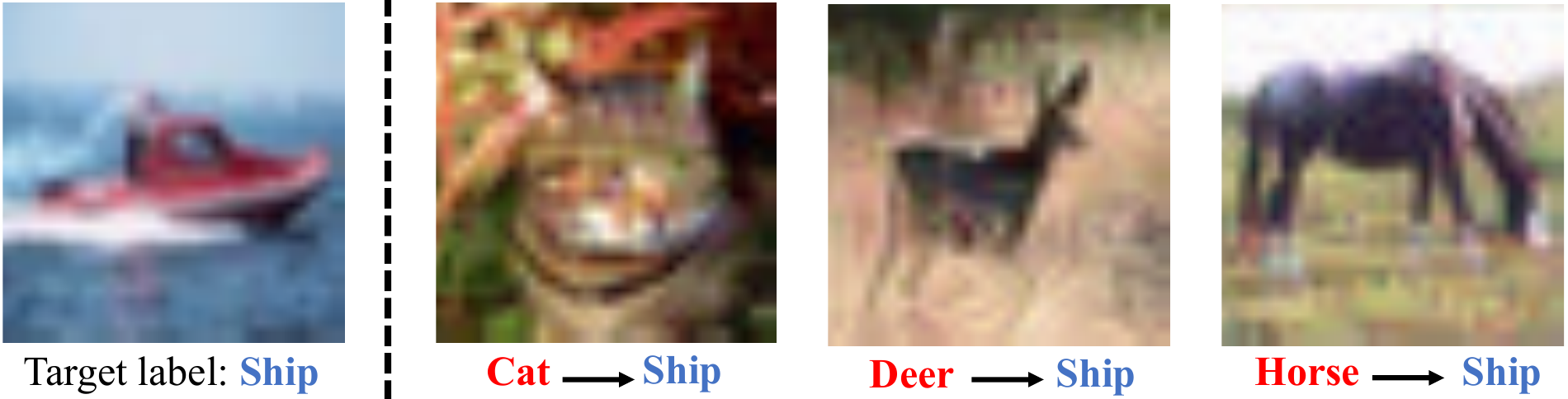}
    \caption{CIFAR-10}
\end{subfigure}    
\begin{subfigure}{1\textwidth}
    \centering
    \includegraphics[width=0.6\textwidth]{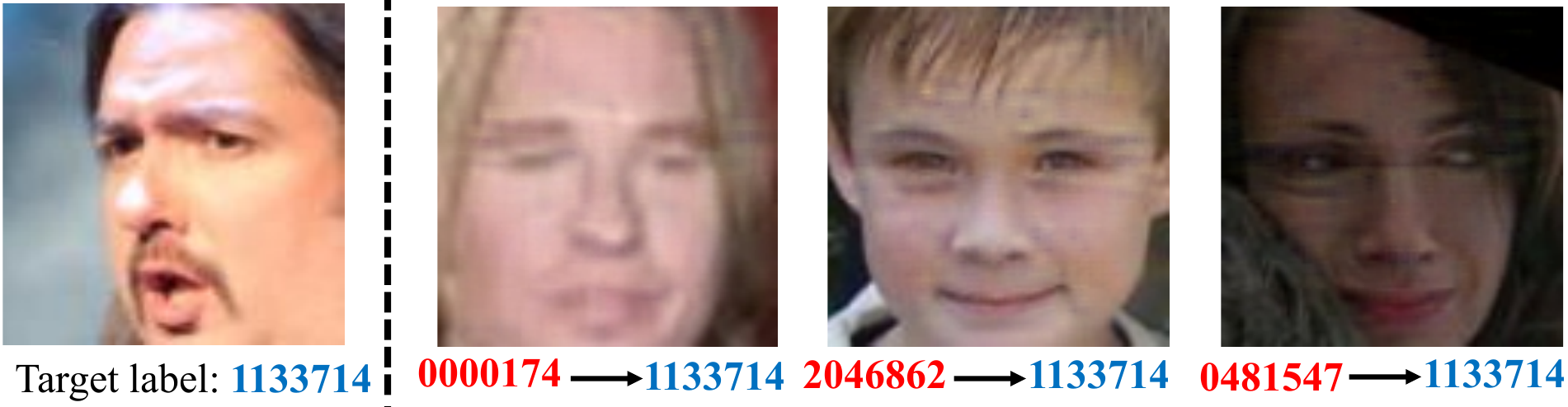}
    \caption{restricted CASIA-WebFace}
\end{subfigure}
\caption{Some examples of the target image and modified image on the CIFAR-10 and restricted CASIA-WebFace dataset. \textbf{The first column}: target images from the original datasets. \textbf{The next three columns}: correspondingly modified images generated by the MDP. The red and blue words indicate the original and the target label, respectively.}
\label{fig:mislabel_images}
\end{center}
\vspace{-1em}
\end{figure*}

\begin{table*}[ht] 
\centering
 \caption{The transferability evaluation on the $\mathcal{D}_{mod-CIFAR10}$ and $\mathcal{D}_{mod-CASIA}$ dataset.}
 \vspace{-0.2em}
 \begin{minipage}[b]{0.37\textwidth} 
       \begin{tabular}{cc} 
       \toprule
        Network Architecture & $\mathcal{D}_{mod-CIFAR10}$ \\
        \hline
        ResNet-50 & 85.52\\
        ResNet-154 & 86.08\\
        DenseNet-154 & 83.15\\ 
        \bottomrule
    \end{tabular}
  \end{minipage}
 \begin{minipage}[b]{0.37\textwidth} 
       \begin{tabular}{cc} 
       \toprule
        Network Architecture & $\mathcal{D}_{mod-CASIA}$\\
        \hline
        IR-50 & 83.04\\
        IR-152 & 84.37\\
        IR-SE-50 & 81.37\\ 
        \bottomrule
    \end{tabular}
  \end{minipage}
\label{tab2}
\vspace{-1em}
\end{table*}

\subsection{Verification on CIFAR-10 and CASIA-WebFace}
\label{ssec:subhead}

In this experiment, we construct two datasets $\mathcal{D}_{mod-CIFAR10}$ and $\mathcal{D}_{mod-CASIA}$ based on CIFAR-10 and restricted CASIA-WebFace, respectively. The left-hand side of Fig.\ref{fig:unstable_train_effect} represents the test accuracy of the model trained on CIFAR-10 and $\mathcal{D}_{mod-CIFAR10}$, while the right-hand side indicates the performance of the model trained on restricted CASIA-WebFace and $\mathcal{D}_{mod-CASIA}$. The result shows that DNNs trained on the modified dataset can generalize well on the benign testing set, and therefore the effectiveness of MDP is verified.

\begin{table}[ht]
\centering
\caption{Test accuracy of MDP and AugMDP on modified dataset constructed based on CIFAR-10 and restricted CASIA-WebFace dataset, respectively.}
\begin{tabular}{ c | c | c }
\toprule
 & MDP & AugMDP (2) \\
\hline
CIFAR-10 & 85.52 & \textbf{89.04} \\ 
restricted CASIA-WebFace & 83.04 & \textbf{84.67} \\ 
\bottomrule
\end{tabular}
\label{table:aug_effect}
\vspace{-1em}
\end{table}

Fig. \ref{fig:mislabel_images} illustrates some target images and their correspondingly modified images. To quantitively assess the similarity between the selected image and corresponding modified image, we also calculate their structural similarity index (SSIM) \cite{SSIM}. The mean SSIM for CIFAR-10 and restricted CASIA-WebFace are 96.9\% and 96.1\%, respectively. The results show that the modified image is highly similar to the selected image, therefore the invisibility of modification is guaranteed.

\subsection{Discussion}
\label{ssec:augmentation_subsection}

\noindent \textbf{Augmentation Effects.}
To verify the effectiveness of the augmentation effects in AugMDP, we compare AugMDP and MDP on both CIFAR-10 and restricted CASIA-WebFace datasets. Table \ref{table:aug_effect} shows the test accuracy of these two methods on two datasets, where the number in the parenthesis following AugMDP in the table is the value of augmentation-related hyper-parameter $T$. Particularly, AugMDP (1) is equivalent to MDP. As demonstrated in Table \ref{table:aug_effect}, AugMDP is better than MDP across different tasks even when $T$ is relatively small ($i.e., T=2$). Note that $T$ should be adjusted according to specific requirements since AugMDP brings additional computation and storage costs.

\vspace{0.4em}
\noindent \textbf{Transferability.}
In this experiment, we verify whether the modified dataset generated by a given network is also effective for training other network architectures. Table \ref{tab2} shows the test accuracy of several architectures (ResNet-50, ResNet-154, and DenseNet-154 \cite{densenet}) trained on $\mathcal{D}_{mod-CIFAR10}$ generated by ResNet-50, and the test accuracy of IR-50, IR-152, and IR-SE-50\footnote{IR-SE-50 combines IR-50 and SENet \cite{senet}.} trained on $\mathcal{D}_{mod-CASIA}$ generated by IR-50. The result shows that the modified dataset is also effective for training different (especially similar) network architectures. It is probably because the unstable features learned by similar classifiers share certain similarities. Detailed reasons will be further explored in our future work.

\section{Conclusions}
\label{sec:print}
In this paper, we propose the mapping distortion based protection (MDP) and its augmentation-based extension (AugMDP) to protect the visual data privacy in classification tasks by modifying the training set. This method is motivated by the fact that DNNs utilize some useful yet unstable features, which can be modified invisibly. Based on this method, we can protect privacy when the dataset is leaked, while still achieve good performance on the benign testing set when the model is trained on the modified dataset. Extensive experiments are conducted, which verify the feasibility and effectiveness of the proposed methods.

\vfill\pagebreak

\label{sec:ref}

\bibliographystyle{IEEEbib}
\bibliography{ms}

\begin{thebibliography}{10}

\bibitem{zhu2019feature}
Chenchen Zhu, Yihui He, and Marios Savvides,
\newblock ``Feature selective anchor-free module for single-shot object
  detection,''
\newblock in {\em CVPR}, 2019.

\bibitem{wang2019}
Xudong Wang, Zhaowei Cai, Dashan Gao, and Nuno Vasconcelos,
\newblock ``Towards universal object detection by domain attention,''
\newblock in {\em CVPR}, 2019.

\bibitem{ahmed2020density}
Syeda~Mariam Ahmed and Chee~Meng Chew,
\newblock ``Density-based clustering for 3d object detection in point clouds,''
\newblock in {\em CVPR}, 2020.

\bibitem{zhang2019ranksrgan}
Wenlong Zhang, Yihao Liu, Chao Dong, and Yu~Qiao,
\newblock ``Ranksrgan: Generative adversarial networks with ranker for image
  super-resolution,''
\newblock in {\em ICCV}, 2019.

\bibitem{dai2019second}
Tao Dai, Jianrui Cai, Yongbing Zhang, Shu-Tao Xia, and Lei Zhang,
\newblock ``Second-order attention network for single image super-resolution,''
\newblock in {\em CVPR}, 2019.

\bibitem{soh2020meta}
Jae~Woong Soh, Sunwoo Cho, and Nam~Ik Cho,
\newblock ``Meta-transfer learning for zero-shot super-resolution,''
\newblock in {\em CVPR}, 2020.

\bibitem{Dai_2019_CVPR}
Kenan Dai, Dong Wang, Huchuan Lu, Chong Sun, and Jianhua Li,
\newblock ``Visual tracking via adaptive spatially-regularized correlation
  filters,''
\newblock in {\em CVPR}, 2019.

\bibitem{voigtlaender2020siam}
Paul Voigtlaender, Jonathon Luiten, Philip~HS Torr, and Bastian Leibe,
\newblock ``Siam r-cnn: Visual tracking by re-detection,''
\newblock in {\em CVPR}, 2020.

\bibitem{liao2020pg}
Bingyan Liao, Chenye Wang, Yayun Wang, Yaonong Wang, and Jun Yin,
\newblock ``Pg-net: Pixel to global matching network for visual tracking,''
\newblock in {\em ECCV}, 2020.

\bibitem{k-anonymity}
Latanya Sweeney,
\newblock ``k-anonymity: A model for protecting privacy,''
\newblock {\em International Journal of Uncertainty, Fuzziness and
  Knowledge-Based Systems}, vol. 10, no. 05, pp. 557--570, 2002.

\bibitem{texture_1}
Robert Geirhos, Patricia Rubisch, Claudio Michaelis, Matthias Bethge, Felix~A
  Wichmann, and Wieland Brendel,
\newblock ``Imagenet-trained cnns are biased towards texture; increasing shape
  bias improves accuracy and robustness,''
\newblock in {\em ICLR}, 2019.

\bibitem{texture_2}
Leon~A Gatys, Alexander~S Ecker, and Matthias Bethge,
\newblock ``Texture and art with deep neural networks,''
\newblock {\em Current Opinion in Neurobiology}, vol. 46, pp. 178--186, 2017.

\bibitem{ilyas2019adversarial}
Andrew Ilyas, Shibani Santurkar, Logan Engstrom, Brandon Tran, and Aleksander
  Madry,
\newblock ``Adversarial examples are not bugs, they are features,''
\newblock in {\em NeurIPS}, 2019.

\bibitem{no_bugs}
Andrew Ilyas, Shibani Santurkar, Dimitris Tsipras, Logan Engstrom, Brandon
  Tran, and Aleksander Madry,
\newblock ``Adversarial examples are not bugs, they are features,''
\newblock in {\em NeurIPS}, 2019.

\bibitem{profile_1}
Matthew~D Zeiler and Rob Fergus,
\newblock ``Visualizing and understanding convolutional networks,''
\newblock in {\em ECCV}, 2014.

\bibitem{profile_2}
Samuel Ritter, David~GT Barrett, Adam Santoro, and Matt~M Botvinick,
\newblock ``Cognitive psychology for deep neural networks: A shape bias case
  study,''
\newblock in {\em ICML}, 2017.

\bibitem{PGD}
Aleksander Madry, Aleksandar Makelov, Ludwig Schmidt, Dimitris Tsipras, and
  Adrian Vladu,
\newblock ``Towards deep learning models resistant to adversarial attacks,''
\newblock in {\em ICLR}, 2018.

\bibitem{CIFAR10}
Alex Krizhevsky et~al.,
\newblock ``Learning multiple layers of features from tiny images,''
\newblock Tech. {R}ep., Citeseer, 2009.

\bibitem{CASIA}
Yi~Dong, Lei Zhen, Shengcai Liao, and Stan~Z. Li,
\newblock ``Learning face representation from scratch,''
\newblock {\em Computer Science}, 2014.

\bibitem{resnet}
Kaiming He, Xiangyu Zhang, Shaoqing Ren, and Jian Sun,
\newblock ``Deep residual learning for image recognition,''
\newblock in {\em CVPR}, 2016.

\bibitem{arcface}
Jiankang Deng, Jia Guo, Niannan Xue, and Stefanos Zafeiriou,
\newblock ``Arcface: Additive angular margin loss for deep face recognition,''
\newblock in {\em CVPR}, 2019.

\bibitem{SSIM}
Zhou Wang, Alan~C Bovik, Hamid~R Sheikh, and Eero~P Simoncelli,
\newblock ``Image quality assessment: from error visibility to structural
  similarity,''
\newblock {\em IEEE transactions on image processing}, vol. 13, no. 4, pp.
  600--612, 2004.

\bibitem{densenet}
Gao Huang, Zhuang Liu, Laurens Van Der~Maaten, and Kilian~Q Weinberger,
\newblock ``Densely connected convolutional networks,''
\newblock in {\em CVPR}, 2017.

\bibitem{senet}
Jie Hu, Li~Shen, and Gang Sun,
\newblock ``Squeeze-and-excitation networks,''
\newblock in {\em CVPR}, 2018.

\end{thebibliography}

\end{document}